\pdfoutput=1

\documentclass[11pt,twoside,a4paper,cmspaper,final,collab]{cms-tdr}
\def\svnVersion{365912}\def\cmsCernNoTag{CERN-PH-EP/2015-310}\def\cmsCernDate{\today}\def\cmsMessage{Published in Physics Letters B as \href{http://dx.doi.org/10.1016/j.physletb.2016.07.065}{\doi{10.1016/j.physletb.2016.07.065.}}}

\begin{document}\cmsNoteHeader{HIN-15-003}

\hyphenation{had-ron-i-za-tion}
\hyphenation{cal-or-i-me-ter}
\hyphenation{de-vices}
\RCS$HeadURL: svn+ssh://svn.cern.ch/reps/tdr2/papers/HIN-15-003/trunk/HIN-15-003.tex $
\RCS$Id: HIN-15-003.tex 352658 2016-07-01 21:42:19Z alverson $
\newlength\cmsFigWidth
\ifthenelse{\boolean{cms@external}}{\setlength\cmsFigWidth{0.49\textwidth}}{\setlength\cmsFigWidth{0.97\textwidth}}
\ifthenelse{\boolean{cms@external}}{\providecommand{\cmsLeft}{top\xspace}}{\providecommand{\cmsLeft}{left\xspace}}
\ifthenelse{\boolean{cms@external}}{\providecommand{\cmsRight}{bottom\xspace}}{\providecommand{\cmsRight}{right\xspace}}

\newcommand{\QQbar}{\ensuremath{\mathrm{Q} \overline{\mathrm{Q}}}\xspace}
\newcommand{\UpsOne}{\ensuremath{\Upsilon\mathrm{(1S)}}\xspace}
\newcommand{\UpsTwo}{\ensuremath{\Upsilon\mathrm{(2S)}}\xspace}
\newcommand{\UpsThree}{\ensuremath{\Upsilon\mathrm{(3S)}}\xspace}
\newcommand{\UpsN}{\ensuremath{\Upsilon\mathrm{(nS)}}\xspace}
\newcommand{\cpm}{\ensuremath{N_\text{ch}}\xspace}
\newcommand{\lth}{\ensuremath{\lambda_{\vartheta}}\xspace}
\newcommand{\lph}{\ensuremath{\lambda_{\varphi}}\xspace}
\newcommand{\ltp}{\ensuremath{\lambda_{\vartheta \varphi}}\xspace}
\newcommand{\ltilde}{\ensuremath{\tilde{\lambda}}\xspace}
\newcommand{\costh}{\ensuremath{\cos\vartheta}\xspace}
\newcommand{\fbg}{\ensuremath{f_\text{Bg}}\xspace}

\cmsNoteHeader{HIN-15-003}

\title{\texorpdfstring{\UpsN polarizations versus particle multiplicity in pp collisions at $\sqrt{s} = 7$\TeV}
{Y(nS) polarizations versus particle multiplicity in pp collisions at sqrt(s) = 7 TeV}}

\date{\today}

\abstract{
The polarizations of the \UpsOne, \UpsTwo, and \UpsThree
mesons are measured as a function of the charged particle multiplicity
in proton-proton collisions at $\sqrt{s} = 7$\TeV.
The measurements are performed with a dimuon data sample collected in 2011
by the CMS experiment, corresponding to an integrated luminosity of 4.9\fbinv.
The results are extracted from the dimuon decay angular distributions,
in two ranges of \UpsN transverse momentum
(10--15 and 15--35\GeV), and in the rapidity interval $\abs{y}<1.2$.
The results do not show significant changes from low- to high-multiplicity pp collisions,
although large uncertainties preclude definite statements
in the \UpsTwo and \UpsThree cases.}

\hypersetup{%
pdfauthor={CMS Collaboration},%
pdftitle={Y(nS) polarizations versus particle multiplicity in pp collisions at sqrt(s) = 7 TeV},%
pdfsubject={CMS},%
pdfkeywords={CMS, physics, quarkonium production, quarkonium polarization, QCD medium effects}}

\maketitle

\section{Introduction}

Studies of heavy-quarkonium production contribute to an improved understanding
of hadron formation within the context of quantum chromodynamics (QCD)~\cite{bib:Brambilla2011}.
Quarkonium production is expected to proceed in two steps~\cite{bib:NRQCD}.
First, a heavy quark-antiquark pair, \QQbar, is produced, with angular momentum $L$ and spin $S$.
Then this ``pre-resonance" binds into the measured quarkonium state through
a nonperturbative evolution that may change $L$ and/or $S$.
The short-distance \QQbar production cross sections (SDCs)
are functions of the \QQbar momentum and
are calculated in perturbative QCD~\cite{bib:BK, bib:GWWZ, bib:Chao, bib:GWWZups},
while the proba\-bilities that \QQbar pairs
of different quantum properties form the observed quarkonium state are parametrized by
momentum-independent long-distance matrix elements (LDMEs).
Since they are expected to scale with powers of the heavy-quark velocity squared, $v^2$,
in the nonrelativistic limit ($v^2 \ll 1$) most LDMEs are negligible and
\mbox{$S$-wave} vector quarkonia should be dominantly formed from \QQbar pairs produced
as colour-singlet, $^3S_1^{[1]}$, or as one of the $^1S_0^{[8]}$, $^3S_1^{[8]}$ and $^3P_J^{[8]}$ colour-octet states.
While the colour-singlet LDME can be calculated with potential models, the others,
reflecting the complexity of the evolution of a coloured QCD system
into a formed hadron, are determined through phenomenological analyses of quarkonium
production data~\cite{bib:BK, bib:GWWZ, bib:Chao, bib:GWWZups, bib:FKLSW}.
Polarization data play a central role in these analyses~\cite{bib:FKLSW},
which are performed in the zero-momentum frame of
the quarkonium state (and, approximately, of the \QQbar pair) and can directly reveal the
quantum properties of the \QQbar, relying in most cases only on basic angular-momentum analysis.
For example, $^1S_0^{[8]}$ \QQbar states evolve into unpolarized $^3S_1$ quarkonia,
while $^3S_1^{[8]}$ states, with quantum numbers identical to those of a gluon,
lead to transversely polarized $^3S_1$ quarkonia.

The factorization hypothesis
of nonrelativistic QCD
implicitly assumes that the LDMEs are universal constants, independent
of the short-distance process that created the \QQbar: the same LDMEs
should be extracted from proton-(anti)proton and $\Pep\Pem$ data, for example.
However, cross section and polarization measurements at high transverse momentum, \pt,
are currently limited to \Pp\Pp\ collisions, so that the LDME universality hypothesis
remains a nontrivial assumption requiring direct experimental investigation.
Since the nonperturbative quarkonium formation process involves interactions with the
QCD medium surrounding the \QQbar state,
allowing it to neutralize its net colour through emission or absorption of soft gluons,
it is important to verify if the polarizations (directly related to the LDMEs)
depend on the complexity of the hadronic environment created by the collision.
Probing if the polarizations are affected by an increase in the
multiplicity of particles produced in \Pp\Pp\ collisions,
the topic of the present analysis, is a first step in such a study,
to be followed by analogous investigations using proton-nucleus and
nucleus-nucleus data collected at different collision centralities.
Such studies are crucial for a reliable interpretation of the quarkonium suppression patterns
seen in high-energy nuclear collisions (see Ref.~\cite{bib:Chatrchyan:2012lxa} and references therein)
and of their relation to signatures of quark-gluon plasma formation~\cite{bib:MatsuiSatz,bib:KarschSatz,bib:Andronic:2015wma}.
While changes in integrated yields or in \pt and rapidity, $y$, distributions can be caused by
effects such as modified parton densities in the nucleus or parton energy loss,
the observation of changes in quarkonium polarization
would be a direct signal of a modification in the bound-state formation mechanism.

This Letter
reports how the polarizations of the \UpsOne, \UpsTwo, and \UpsThree mesons produced
in \Pp\Pp\ collisions at a centre-of-mass energy of 7\TeV
change as a function of charged particle multiplicity, \cpm.
It complements two observations made for \Pp\Pp\ and \Pp Pb collisions~\cite{bib:JHEP04}:
the \UpsN cross sections, normalised by their \cpm-integrated values, increase with \cpm;
the \UpsTwo and \UpsThree cross sections, normalised by the \UpsOne value, decrease with \cpm.

The measurements are performed using a dimuon data sample collected in 2011
by the CMS experiment at the CERN LHC, corresponding to an integrated luminosity of 4.9\fbinv,
and follow the analysis method used in the \cpm-integrated measurement~\cite{bib:CMS-upsilon-pol}.
The dimuon mass distribution is used to separate the \UpsN signals from each other
and from muon pairs due to other processes, such as
decays of heavy flavor mesons.
The \UpsN polarizations are characterized through three parameters,
$\vec{\lambda} = (\lth, \lph, \ltp)$,
reflecting the anisotropy of the angular distribution of the decay muons~\cite{bib:Faccioli-LTVio},
\begin{linenomath}
\begin{equation}
\label{eq:angular_distribution}
\ifthenelse{\boolean{cms@external}}{
\begin{split}
& W(\costh,\varphi|\vec{\lambda}) \, \propto \,
\frac{1}{(3 + \lambda_{\vartheta})}(1  +  \lambda_{\vartheta} \cos^2 \vartheta\, + \\
&
+ \lambda_{\varphi} \sin^2 \vartheta \cos 2 \varphi
+ \lambda_{\vartheta \varphi} \sin 2 \vartheta \cos \varphi ),
\end{split}
}{%
W(\costh,\varphi|\vec{\lambda}) \, \propto \,
\frac{1}{(3 + \lambda_{\vartheta})}(1  +  \lambda_{\vartheta} \cos^2 \vartheta\,
+ \lambda_{\varphi} \sin^2 \vartheta \cos 2 \varphi
+ \lambda_{\vartheta \varphi} \sin 2 \vartheta \cos \varphi ),
}
\end{equation}
\end{linenomath}
where $\vartheta$ and $\varphi$ are the polar and azimuthal angles, respectively, of the
$\Pgmp$.
These $\vec{\lambda}$ parameters, as well as the frame-invariant parameter
$\tilde{\lambda} = (\lambda_\vartheta + 3 \, \lambda_\varphi) /
(1-\lambda_\varphi)$~\cite{bib:Faccioli-PRL-FrameInv},
are measured in
the centre-of-mass helicity frame (HX),
where the $z$ axis coincides with the direction of the $\Upsilon$ momentum.
The $y$ axis of the polarization frame is reversed between positive and negative rapidity,
a definition that avoids the cancellation of $\ltp$ when integrating events over a symmetrical
range in rapidity. This is explained in Ref.~\cite{bib:Faccioli-EPJC}, which provides a
pedagogical introduction to quarkonium polarization physics.
As in the previous CMS quarkonium polarization measurements~\cite{bib:CMS-upsilon-pol,bib:CMS-psi-pol},
the analysis is exclusively based on measured data:
the 3-momentum vectors of the two muons
(containing the spin alignment information of the decaying \UpsN mesons)
and the muon detection efficiencies.

\section{CMS detector and data analysis}

The CMS apparatus is based on
a superconducting solenoid of 6\unit{m} internal diameter, providing a 3.8\unit{T} field.
Within the solenoid volume are a silicon pixel and strip tracker, a lead tungstate crystal
electromagnetic calorimeter, and a brass and scintillator hadron calorimeter.
Muons are measured with
drift tubes, cathode strip chambers, and resistive-plate chambers.
The main detectors used in this analysis are the silicon tracker and the muon system, which
enable the measurement of muon momenta over the pseudorapidity range $\abs{\eta} < 2.4$.
A more detailed description of the CMS detector, together with a definition of the coordinate system
used and the relevant kinematic variables, can be found in Ref.~\cite{bib:Chatrchyan:2008zzk}.
The events were collected using a two-level trigger system.
The first level uses custom hardware processors to select events with two muons.
The high-level trigger, adding information from the silicon tracker,
selects opposite-sign muon pairs of invariant mass $8.5<M<11.5$\GeV, $\abs{y} < 1.25$ and
$\pt > 5$ or 7\GeV (depending on the instantaneous luminosity);
the dimuon vertex fit $\chi^2$ probability must exceed 0.5\% and
the two muons must have a distance of closest approach smaller than 5\unit{mm}.
Although the trigger logic does not reject events on the basis of the \pt of the single muons,
at mid-rapidity the bending induced by the magnetic field prevents muons of \pt smaller than
$\sim$3\GeV from reaching the muon stations.

\begin{figure}[tbh]
\centering
\includegraphics[width=0.48\textwidth]{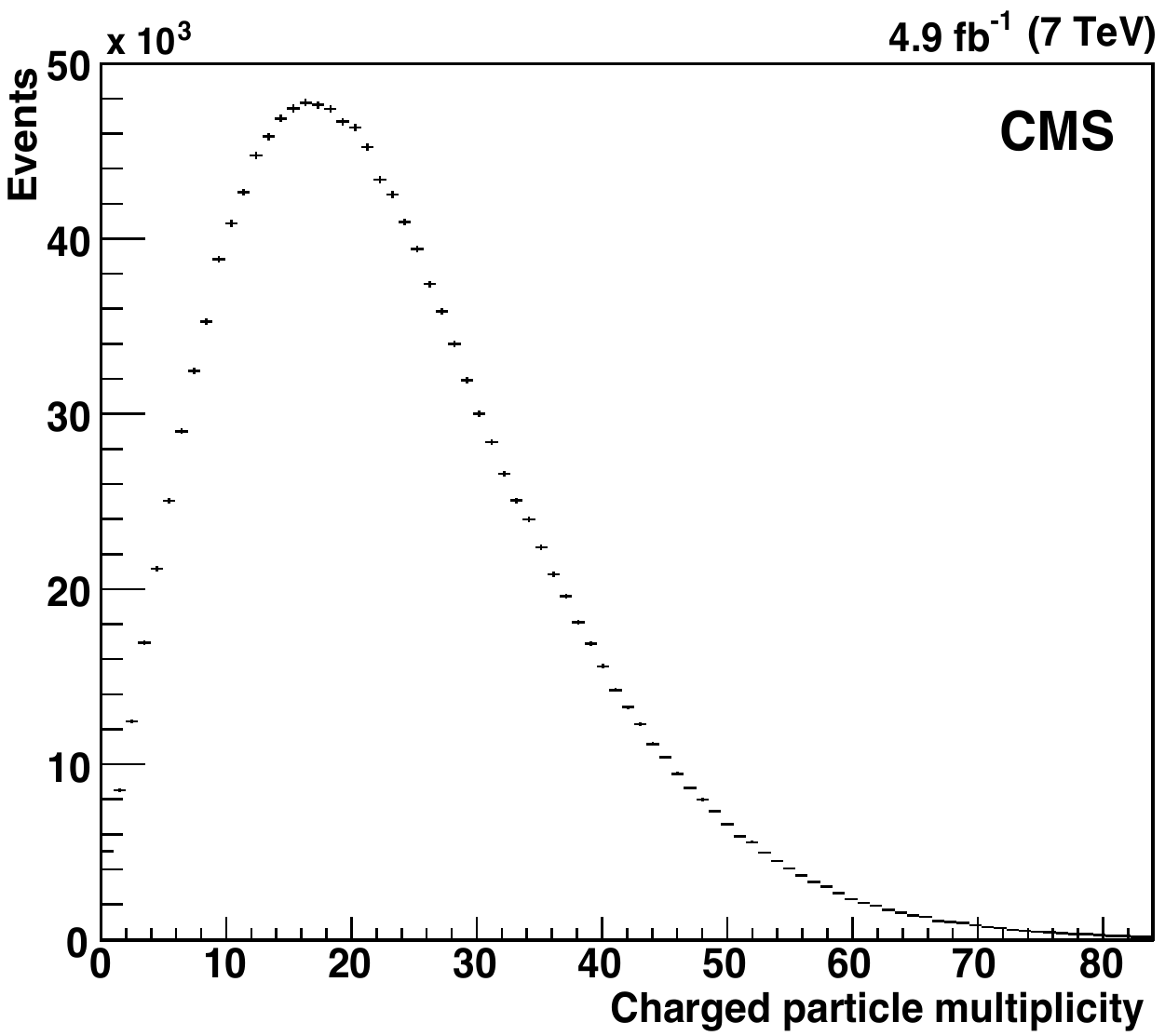}
\caption{Charged particle multiplicity distribution of the events selected for the analysis.}
\label{fig:CPMDistro}
\includegraphics[width=0.48\textwidth]{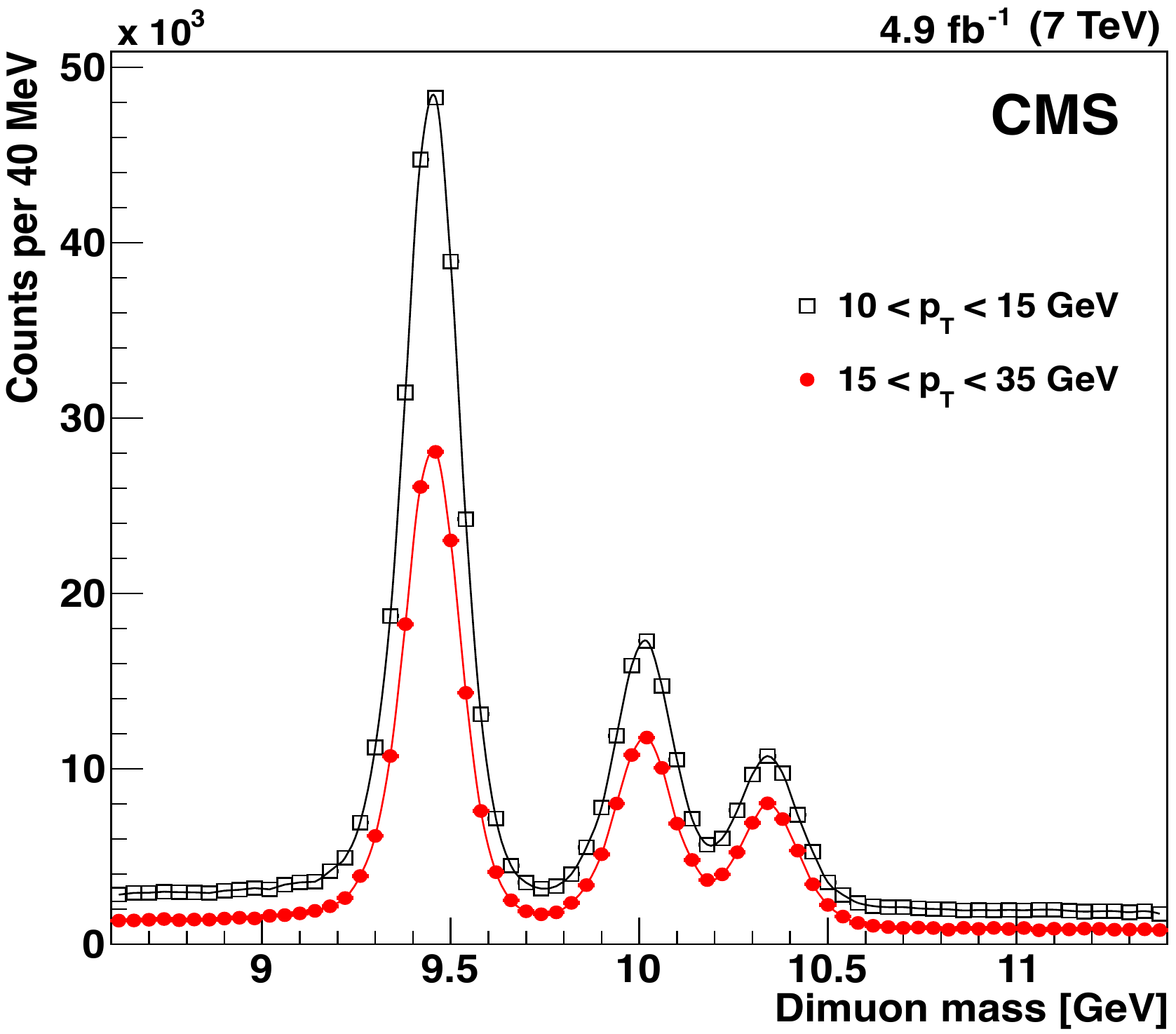}
\caption{Dimuon mass distributions in the \UpsN region for two \pt ranges.}
\label{fig:dimuDist}
\end{figure}

The offline analysis selects muon tracks with hits in more than ten tracker layers,
at least two of which are in the pixel layers, and matched with segments in the muon system.
They must have a good track fit quality, point to the interaction region, and match
the muon objects that triggered the event.
The selected muons are required to satisfy $\abs{\eta} < 1.6$ and to have \pt above 4.5, 3.5, and
3\GeV for $\abs{\eta} < 1.2$, $1.2 < \abs{\eta} < 1.4$, and $1.4 < \abs{\eta} < 1.6$, respectively, to ensure
reliable detection and trigger efficiencies.
The combinatorial background from uncorrelated muons is suppressed by requiring a dimuon
vertex fit $\chi^2$ probability larger than 1\%
and by rejecting events where the distance between the dimuon vertex and the primary vertex is larger than twice its resolution.
In events with multiple reconstructed primary vertices (pileup), the one
nearest to the point of closest approach between the trajectory of the dimuon and the beam line is selected.
The \cpm variable is computed by counting ``high purity"~\cite{bib:TRK-11-001} charged tracks,
excluding the two muons, of
$\abs{\eta} < 2.4$, $\pt > 500$\MeV,
and \pt measured with better than 10\% relative accuracy.
Acceptance and reconstruction efficiencies are not corrected for.
Each track is assigned a weight reflecting the likelihood that it belongs to the primary vertex~\cite{bib:TRK-11-001};
tracks consistent with the vertex have a weight close to unity.
The migration of events from one \cpm bin to the next, caused by inadvertently counting
spurious tracks produced in near-by pileup vertices, is kept negligible by rejecting
events with more than 16 vertices.
Figure~\ref{fig:CPMDistro} shows the \cpm distribution of the events selected in this analysis.

The dimuon mass distribution, shown in Fig.~\ref{fig:dimuDist},
is well described by three Crystal-Ball functions~\cite{bib:CrystalBall}, one per \UpsN peak,
and a second-order polynomial function representing the underlying continuum,
determined from the mass sidebands, 8.6--8.9 and 10.6--11.4\GeV.
The dimuon mass resolution is $\sigma \approx 80$\MeV, slightly dependent on \pt.
The \UpsN signal mass regions are defined as the ${\pm}1\,\sigma$
windows around the
fitted means of the Crystal-Ball functions.
The corresponding cross-feed between the three peaks is negligible.
The analysis is performed in five \cpm bins, 0--10, 10--20, 20--30, 30--40, and 40--60,
sufficiently numerous and narrow to probe potential variations of the polarizations, and
in two \UpsN \pt ranges,
10--15 and 15--35\GeV.
The dimuons
are integrated within $\abs{y} < 1.2$.
The lower \pt \UpsThree polarization measurement merges the two highest \cpm bins,
to reduce the background-related systematic uncertainties.
In the lowest \cpm bin, the background fractions
in the signal mass regions,
\fbg, are approximately 3\%, 7\%, and 10\%
for the \UpsOne, \UpsTwo, and \UpsThree, respectively. The corresponding values in the
highest \cpm bin are
$\sim$4 and $\sim$2.5 times higher
in the  10--15 and 15--35\GeV \pt ranges, respectively.
All analysis bins have signal yields sufficiently high for a reliable measurement,
the worst case being the 2300 \UpsThree events in the highest \cpm bin at high \pt.
All signal yields and background fractions are tabulated in
\ifthenelse{\boolean{cms@external}}{the supplemental material}{Appendix~\ref{app:supp-mat}}.

The single-muon detection efficiencies are measured with a ``tag-and-probe"
technique~\cite{bib:Khachatryan:2010xn},
using event samples collected with triggers specifically designed for this purpose,
including a sample enriched in dimuons from \JPsi decays
where a muon is combined with another track and the pair is required to have an
invariant mass within
2.8--3.4\GeV.
The procedure was validated with detailed Monte Carlo simulation studies.
The measured efficiencies are parametrized as a function of muon \pt, in eight $\abs{\eta}$ bins.
Their uncertainties, $\sim$2--3\%, reflecting the statistical precision of the calibration
samples and possible imperfections of the parametrization,
are independent of \cpm and identical for the three \UpsN states.
These \emph{global uncertainties} do not affect the search for potential
variations of the polarizations from low- to high-multiplicity events.
The trigger and the selection criteria could potentially
introduce differences between the dimuon detection efficiencies and
the product of the efficiencies of the two single muons.
Simulation studies reveal that
such correlations have a negligible dependence on $\costh$ and $\varphi$,
in the phase space of this analysis~\cite{bib:CMS-upsilon-pol}.
The residual angular dependences are accounted for in the evaluation of
the global systematic uncertainties.

\section{Extraction of the polarization parameters}

The two-dimensional angular distribution, in $\costh$ and $\varphi$,
of the background corresponding to a given \UpsN state
is evaluated as a weighted average of the distributions measured in the two mass sidebands,
the weights reflecting (linearly) the differences between the \UpsN mass
and the median masses of the sidebands.
The background component is subtracted on an event-by-event basis using a likelihood-ratio criterium,
randomly selecting and removing a fraction \fbg of events distributed according to the
$(\pt, \abs{y}, M, \costh, \varphi)$ distribution of the background model~\cite{bib:CMS-upsilon-pol}.
The posterior probability density (PPD) for the average values of the \UpsN polarization
parameters ($\vec{\lambda}$) inside a particular bin is then defined as a product
over the remaining (signal-like) events $i$,
\begin{linenomath}\begin{equation}
\label{eq:total_likelihood}
  \mathcal{P}(\vec{\lambda})= \prod_{i} \mathcal{E}(\vec{p}_1^{\,(i)},\vec{p}_2^{\,(i)}),
\end{equation}\end{linenomath}
where $\mathcal{E}$ represents the event probability distribution as a function of the muon
momenta $\vec{p}_{1,2}$ in event $i$.
This analysis method does not use model-dependent $(\costh, \varphi)$ acceptance maps;
each event is attributed a probability reflecting the full event kinematics
(not only $\costh$ and $\varphi$) and the values of the polarization parameters,
\begin{linenomath}\begin{equation}
\label{eq:event_probability_definition}
\mathcal{E}(\vec{p}_1,\vec{p}_2)= \frac{1}{\mathcal{N}(\vec{\lambda})} \,
W(\costh,\varphi|\vec{\lambda}) \, \epsilon(\vec{p}_1,\vec{p}_2),
\end{equation}\end{linenomath}
where $\epsilon(\vec{p}_1,\vec{p}_2)$ is the measured detection efficiency.
The normalization factor $\mathcal{N}(\vec{\lambda})$ is calculated by integrating
$W \cdot \epsilon$ over $\costh$ and $\varphi$ uniformly,
using $(\pt, \abs{y}, M)$ distributions determined from the background-subtracted data.
To account for the statistical fluctuations related to its random nature,
the background subtraction procedure is repeated 50 times.

\begin{figure}[htbp]
\centering
\includegraphics[width= 0.48\textwidth]{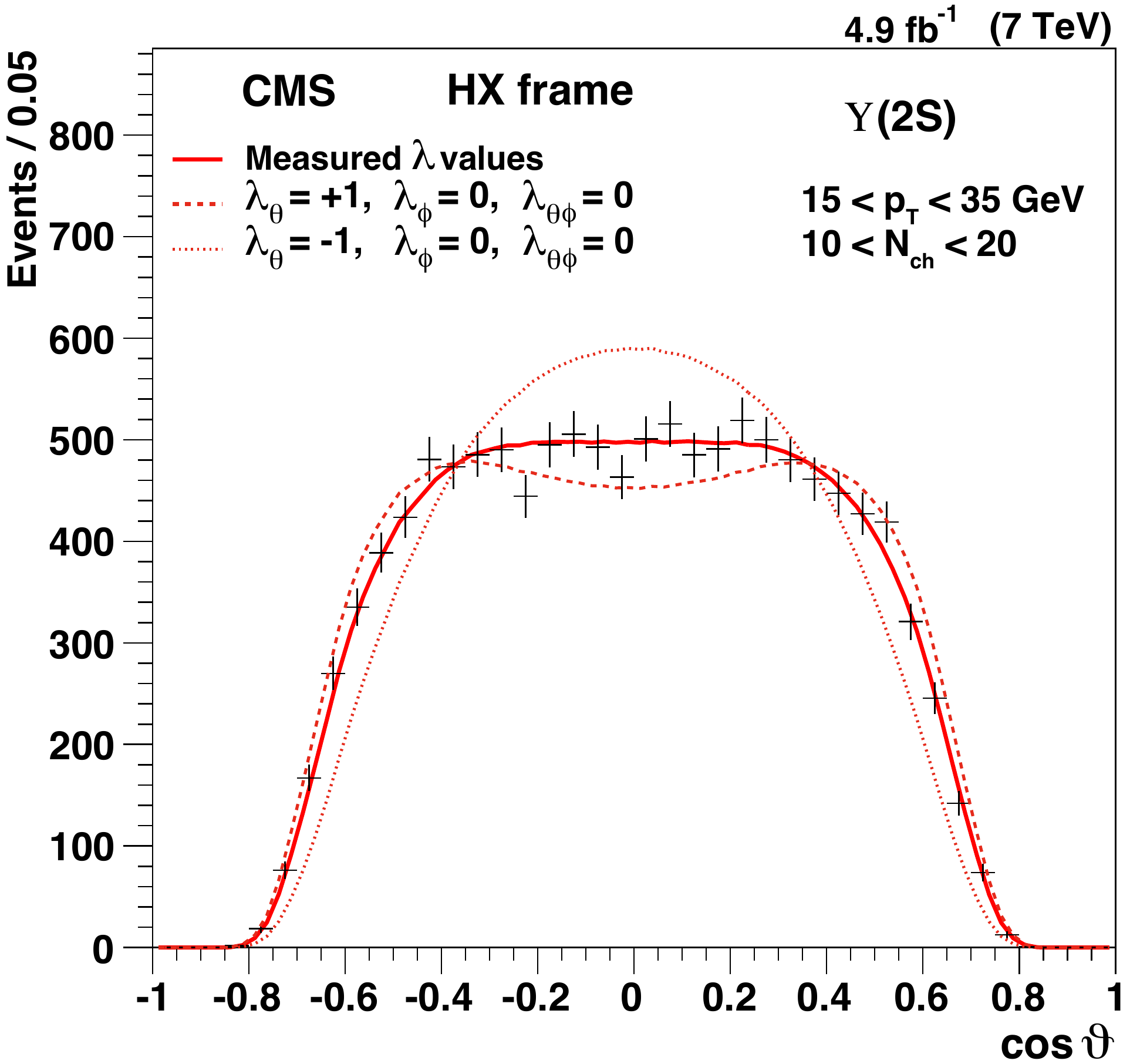}
\includegraphics[width= 0.48\textwidth]{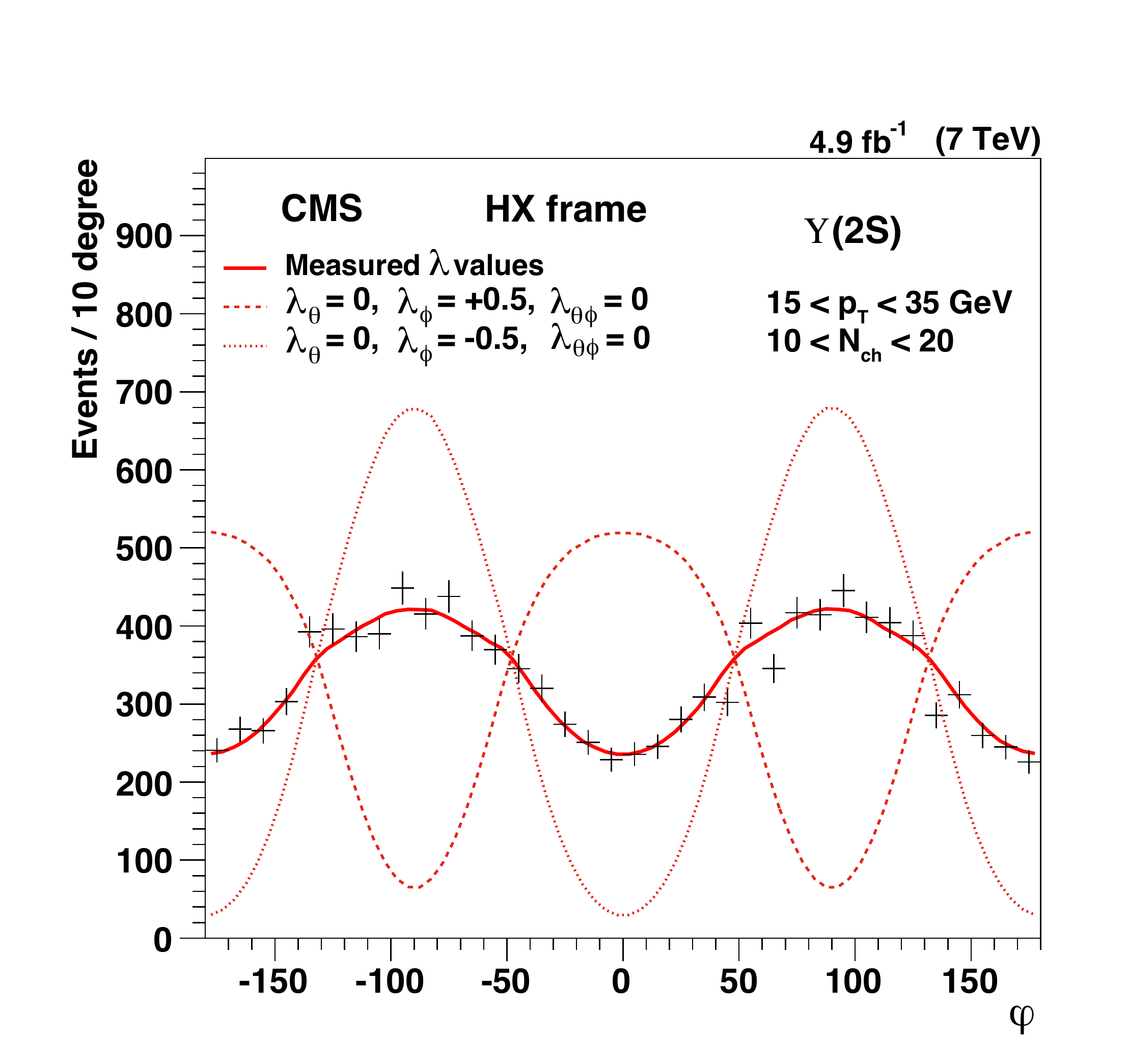}
\caption{Distributions of $\costh$ (\cmsLeft) and $\varphi$ (\cmsRight),
for the \UpsTwo in a representative analysis bin.
The curves represent two polarization scenarios
(dashed and dotted lines, defined in the legends) and the measured case
(solid lines: $\lth=0.237$, $\lph=-0.027$, $\ltp=-0.025$).}
\label{fig:angularProjections}
\end{figure}

Figure~\ref{fig:angularProjections} compares the $\costh$ and $\varphi$ distributions
measured for \UpsTwo signal events of $15<\pt<35$\GeV and $10<\cpm<20$
with curves representing the ``best fit".
For illustration, curves reflecting extreme polarization scenarios are also shown:
fully transverse ($\lth=+1$) and fully longitudinal ($\lth=-1$) in the $\costh$ panel,
and $\lph=\pm 0.5$ in the $\varphi$ panel ($\abs{\lph}$ must be smaller than 0.5 if $\lth=0$~\cite{bib:Faccioli-LTVio}).

Each of the systematic uncertainties on the polarization parameters
caused by the analysis framework and the detection efficiencies
is individually evaluated through 50 statistically independent pseudo-experiments.
For each effect, the systematic uncertainty is the difference between the
injected and resulting parameters.
The robustness of the framework to measure the signal polarization is validated for several
signal and background polarization scenarios. The impact of residual biases that could be
caused by uncertainties on the muon or dimuon efficiencies is evaluated by extracting the
polarization parameters after applying corresponding variations to the input efficiencies.
The background model uncertainty is evaluated by modifying
the relative weights
of the low- and high-mass sidebands when building the background distributions.
A broad range of hypotheses is considered, including the assumption that the background under
the \UpsOne (\UpsThree) peak resembles exclusively the low-mass (high-mass) sideband,
or assuming that it is reproduced by an equal mixture of the two sideband distributions.
Several systematic uncertainties have similar levels,
except in the highest \cpm bins and the lowest \pt range,
where the background model uncertainty dominates,
especially for the \UpsTwo and \UpsThree states.
For the \UpsOne state and in the HX frame,
the \cpm-dependent systematic uncertainties are
$\sim$0.1 for \lth and $\sim$0.03--0.05 for \lph and \ltp,
slightly increasing with \cpm.
The corresponding \UpsTwo and \UpsThree values are slightly larger:
$\sim$0.2 for \lth, $\sim$0.04 for \lph, and $\sim$0.05--0.08 for \ltp.
The statistical uncertainties are negligible for the \UpsOne state and become
dominant for the \UpsTwo and \UpsThree states, as \cpm increases.

\section{Results}

\begin{figure}[htbp]
\centering
\includegraphics[width=0.49\textwidth]{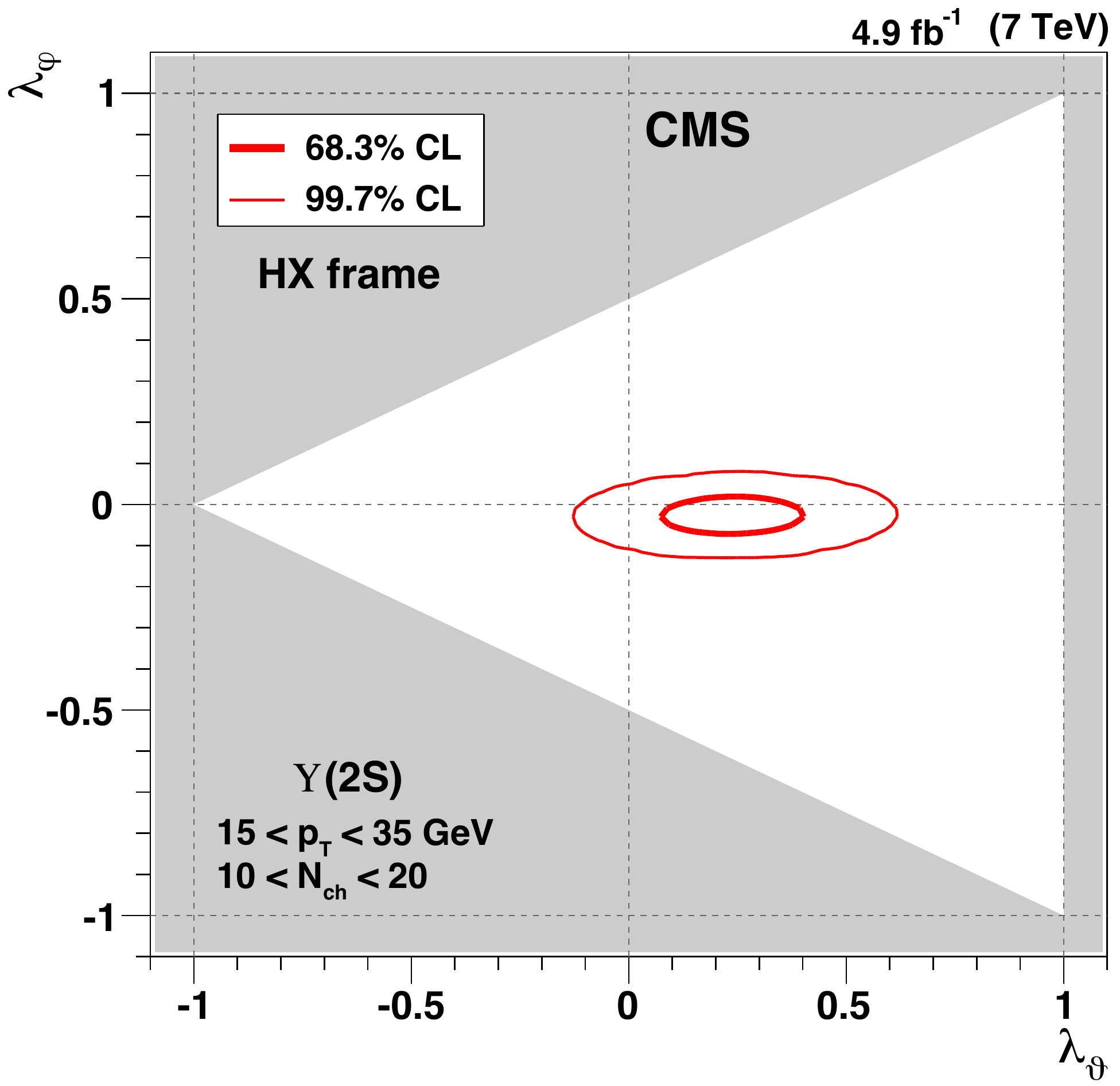}
\includegraphics[width=0.49\textwidth]{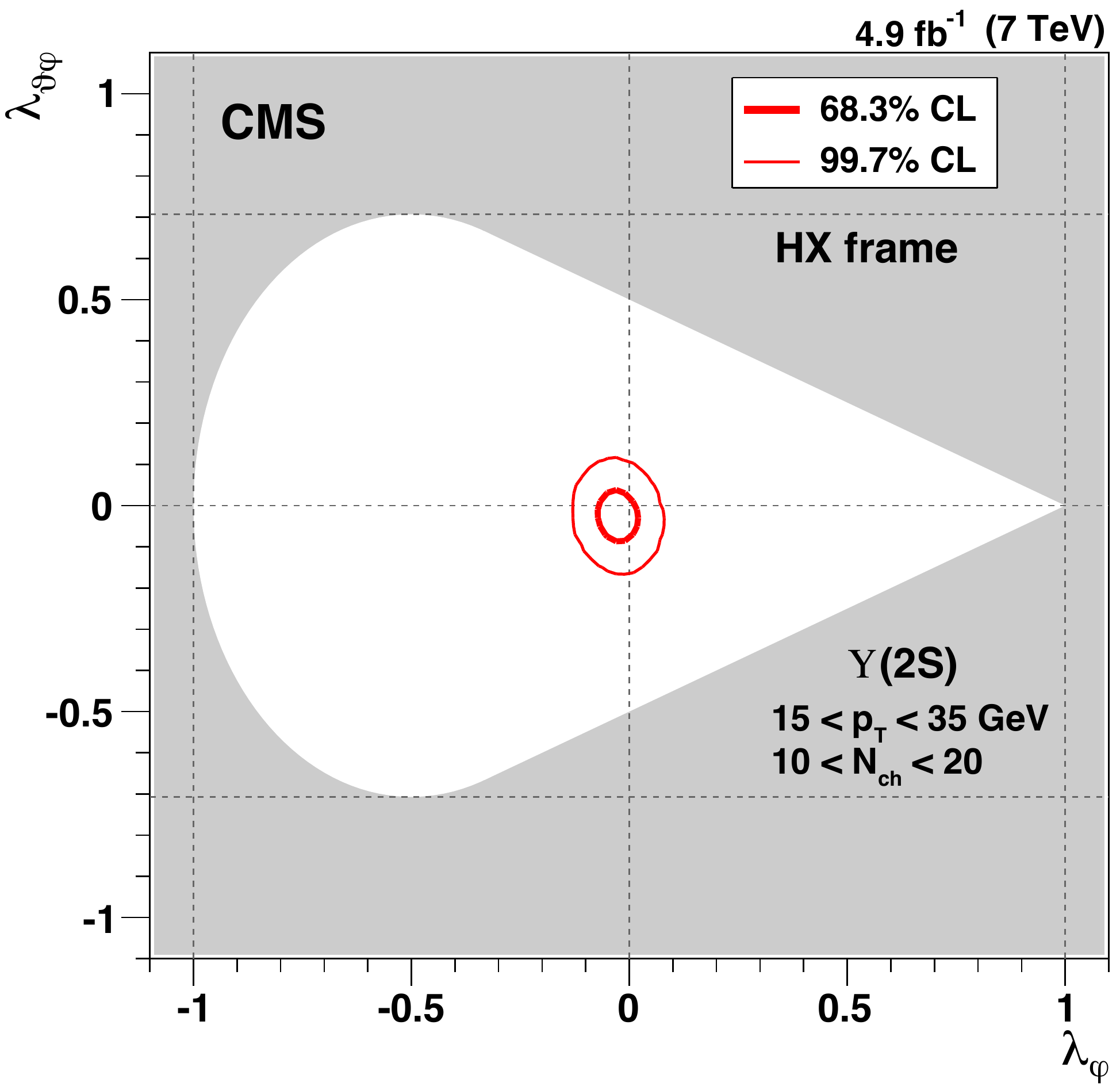}
\caption{Two-dimensional marginals of the PPD for the HX frame
in the \lph vs.\ \lth (\cmsLeft) and \ltp vs.\ \lph (\cmsRight) planes,
for \UpsTwo with $15<\pt<35$\GeV and $10<\cpm<20$,
displaying the 68.3\% and 99.7\% CL total uncertainties.
The shaded areas represent physically forbidden regions of parameter space
for the decay of a $J=1$ particle~\cite{bib:Faccioli-LTVio}.}
\label{fig:2Dcorrelations}
\end{figure}

The final PPD of the polarization parameters is an envelope of the PPDs corresponding
to all hypotheses considered in the evaluation of the systematic uncertainties.
In each analysis bin, the central values and 68.3\% confidence level (CL)
uncertainties of each polarization parameter are evaluated from the
corresponding one-dimensional marginal posterior, calculated by numerical integration.
In the HX frame, the $\lambda$ parameters are measured with negligible correlations,
as illustrated by Fig.~\ref{fig:2Dcorrelations}, which shows the two-dimensional marginals of the PPD
in the \lph vs.\ \lth\ and \ltp vs.\ \lph planes, for a representative analysis bin.

\begin{figure*}[t!bhp]
\centering
\includegraphics[width=0.97\textwidth]{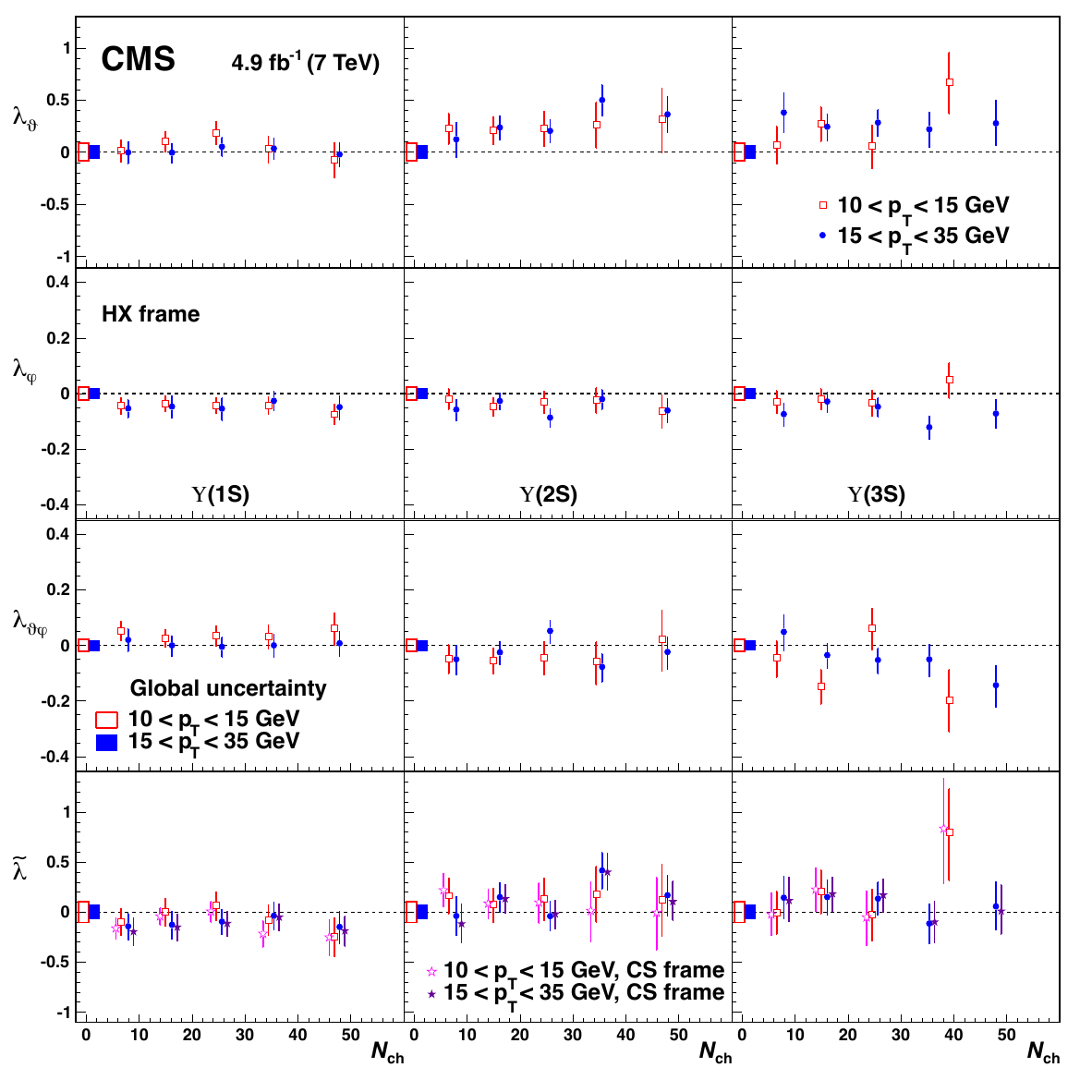}
\caption{The \lth, \lph, \ltp, and \ltilde parameters (top to bottom) for
the \UpsOne, \UpsTwo, and \UpsThree states (left to right), in the HX frame,
as a function of \cpm, for both \pt ranges.
The \ltilde values are also shown for the CS frame;
the HX and CS uncertainties are strongly correlated.
The vertical bars represent the \cpm-dependent total uncertainties
(at 68.3\% CL),
while the boxes at the zero horizontal line represent the global uncertainties.
The points are placed at the average \cpm of each bin, with a small offset for easier viewing.}
\label{fig:ResultsHX}
\end{figure*}

Figure~\ref{fig:ResultsHX} shows the \lth, \lph, \ltp, and \ltilde values
measured in the HX frame for the three \UpsN states, in both \pt ranges.
The corresponding numerical results are tabulated in
\ifthenelse{\boolean{cms@external}}{the supplemental material}{Appendix~\ref{app:supp-mat}}.
The \ltilde values have also been measured in
the Collins--Soper frame (CS)~\cite{bib:CS},
whose $z$ axis is the average of the two beam directions in the $\Upsilon$ rest frame,
and in the perpendicular helicity frame (PX)~\cite{bib:Braaten:2008mz}, orthogonal to the CS frame.
The three measurements agree with each other, within systematic uncertainties (similar in all frames),
as required in the absence of unaccounted systematic effects~\cite{bib:ImprovedQQbarPol}.

Regarding the \UpsOne results, all the $\lambda$ parameters are close to zero,
indicating essentially unpolarized production, as expected if the mesons
included in this analysis would be dominantly produced through the unpolarized
$^1S_0^{[8]}$ pre-resonant octet state. The trend as a function of \cpm does not indicate
any strong changes in \UpsOne production between low- and high-multiplicity \Pp\Pp\ collisions.
The measurements are compatible with a non-negligible fraction of
\UpsTwo and \UpsThree mesons being produced via the transversely
polarized $^3S_1^{[8]}$ octet term. Given the present uncertainties,
no clear trends can be seen regarding changes of their polarizations with \cpm.

\begin{figure}[htb]
\centering
\includegraphics[width=0.49\textwidth]{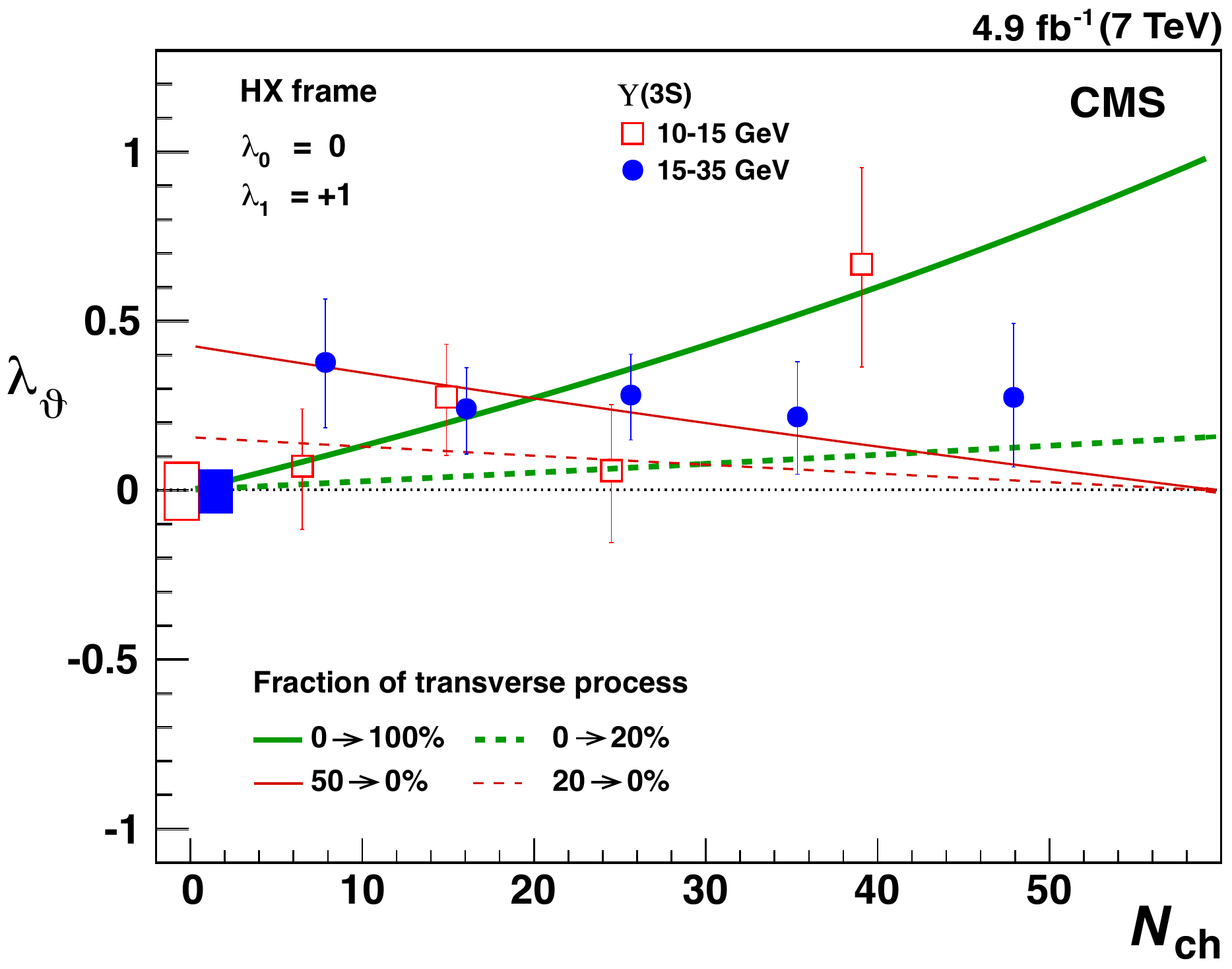}
\includegraphics[width=0.49\textwidth]{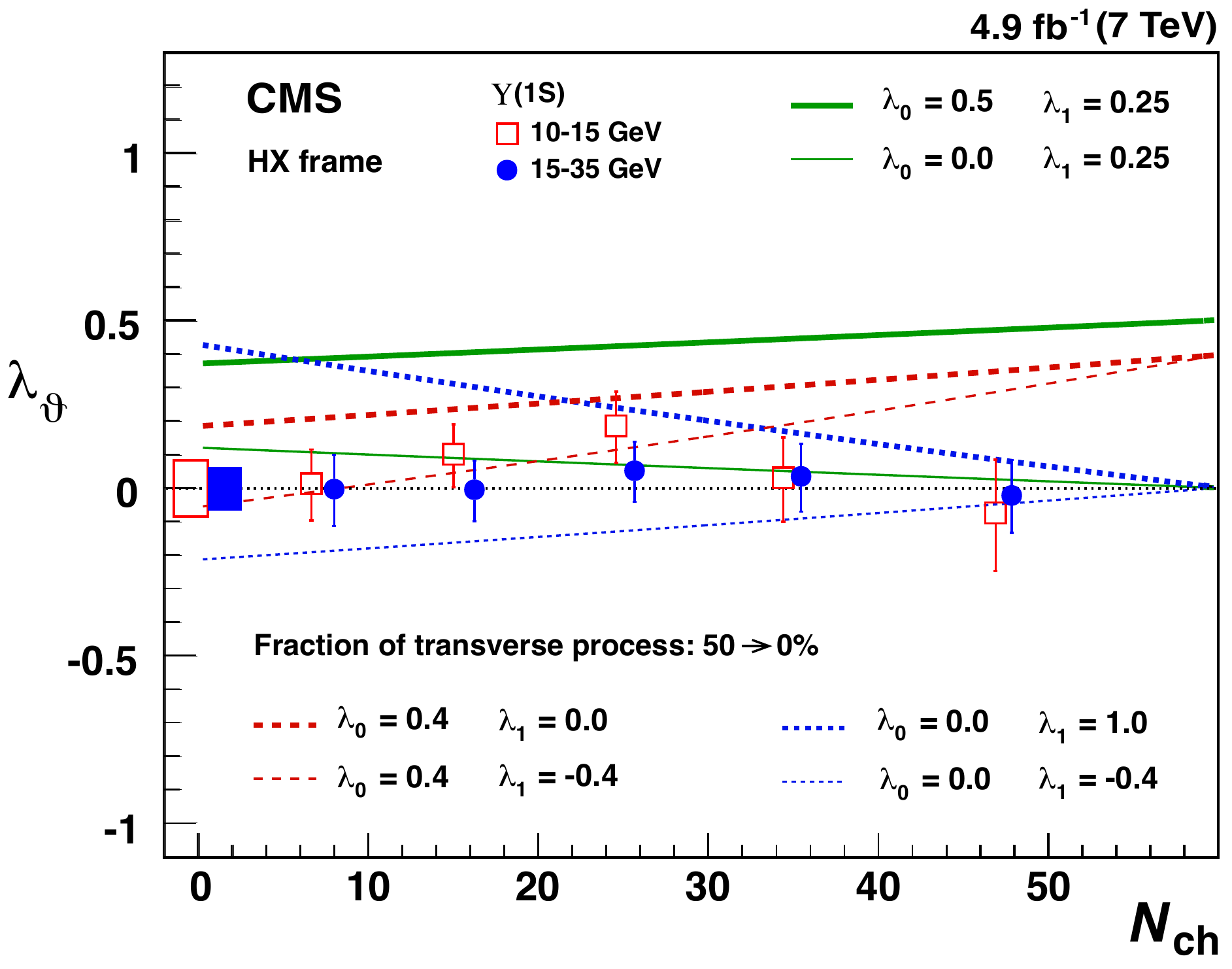}
\caption{Expected \cpm-dependences of the \lth parameter for the sum of two processes,
of polarizations $\lambda_0$ and $\lambda_1$,
of relative fractions changing linearly with \cpm (see text for details).
The measured \lth values are also shown, for the \UpsThree (left) and \UpsOne (right).}
\label{fig:scenarios}
\end{figure}

To place these results into context, Fig.~\ref{fig:scenarios}-\cmsLeft illustrates how the \lth parameter
would change as a function of \cpm if quarkonium production would be dominated by two processes,
one unpolarized ($\lth = 0$, as is the case for the $^1S_0^{[8]}$ octet)
and the other fully transversely polarized in the HX frame
($\lth = +1$, as for the $^3S_1^{[8]}$ octet, at high enough \pt).
The four curves represent different variations with \cpm (linearly in the $0<\cpm<60$ range)
of the fraction of events, $f$, produced through the latter process (defined in the legends).
These curves were computed knowing that
the polarization of a sample of quarkonium states produced through two different processes,
of polarizations $\lambda_0$ and $\lambda_1$, depends on $f$ as~\cite{bib:Faccioli_chi}
\begin{linenomath}
\begin{equation}
\lambda(f) =
\bigg[
\frac{(1-f) \, \lambda_0} {3+\lambda_0} +
\frac{f \, \lambda_1} {3+\lambda_1}
\bigg]
\bigg/\bigg[
\frac{1-f} {3+\lambda_0} +
\frac{f} {3+\lambda_1}
\bigg].
\end{equation}
\end{linenomath}

Changes in the LDMEs,
in particular of the dominant $^1S_0^{[8]}$ and $^3S_1^{[8]}$ octet terms~\cite{bib:FKLSW},
are not the only possible cause of variations in the measured \UpsN $\lambda$ parameters
between low- and high-multiplicity \Pp\Pp\ collisions;
the effects of feed-down decays from heavier quarkonia should also be considered.
In fact, the polarizations reported here correspond to inclusive \UpsN samples,
not distinguishing mesons emitted in the decays of S- and P-wave
bottomonium states from the directly-produced ones.
Assuming that all directly-produced S-wave states have identical polarizations,
their decays to lighter S-wave states
do not induce differences between the measured (inclusive) polarizations
and those of the directly-produced mesons.
On the contrary, feed-down decays from P-wave states can significantly
affect the measured values, especially for the \UpsOne state,
presumably the one affected by the largest feed-down fraction.
It is presently not possible to reliably evaluate the influence of the feed-down
decays on the measured \UpsN polarizations, for lack of information regarding the
$\chi_b$ polarizations and their feed-down fractions.
Figure~\ref{fig:scenarios}-\cmsRight shows how the measured (inclusive) polarization
is expected to change as a function of \cpm if
the directly-produced component (of polarization $\lambda_0$)
is complemented by a feed-down component (of polarization $\lambda_1$)
that contributes with a fraction $f$, decreasing linearly with \cpm from 50\% to 0 in the $0<\cpm<60$ range.
The six curves correspond to different assumptions for $\lambda_0$
and $\lambda_1$, reported in the legends,
with $\lambda_1$ representing an effective average
of the $\chi_{b1}$ and $\chi_{b2}$ polarizations
(the $\chi_{b1}$ and $\chi_{b2}$ \lth values must verify $\lth > -1/3$ and $\lth > -3/5$, respectively~\cite{bib:Faccioli_chi}).
In these scenarios the feed-down fraction is assumed to become
negligible at high \cpm,
where the inclusive \lth tends to the direct $\lambda_0$ value.
At low \cpm, where the feed-down contribution is, hypothetically, the highest,
the inclusive \lth parameter crucially depends on the assumed $\chi_b$ polarization.

\section{Summary}

The polarizations of the \UpsOne, \UpsTwo, and \UpsThree mesons produced in \Pp\Pp\ collisions
at $\sqrt{s} = 7$\TeV have been determined as functions of the
charged particle multiplicity of the event
in two \UpsN \pt ranges.
The measurements do not show significant variations as a function of \cpm,
even though the large \UpsTwo and \UpsThree uncertainties preclude definite statements
in these cases.
This study opens the way for analogous measurements extending to the charmonium family,
particularly interesting for the $\psi$(2S), which is unaffected by feed-down decays and,
therefore, provides a more direct probe of LDME universality.
Equivalent analyses should also be performed in \Pp Pb and PbPb event samples,
in view of evaluating how quark-antiquark bound-state formation is influenced by the
surrounding medium, which is an essential input for the interpretation of quarkonium
suppression patterns in nuclear collisions.

\begin{acknowledgments}
 We congratulate our colleagues in the CERN accelerator departments for the excellent performance of the LHC and thank the technical and administrative staffs at CERN and at other CMS institutes for their contributions to the success of the CMS effort. In addition, we gratefully acknowledge the computing centres and personnel of the Worldwide LHC Computing Grid for delivering so effectively the computing infrastructure essential to our analyses. Finally, we acknowledge the enduring support for the construction and operation of the LHC and the CMS detector provided by the following funding agencies: BMWFW and FWF (Austria); FNRS and FWO (Belgium); CNPq, CAPES, FAPERJ, and FAPESP (Brazil); MES (Bulgaria); CERN; CAS, MoST, and NSFC (China); COLCIENCIAS (Colombia); MSES and CSF (Croatia); RPF (Cyprus); MoER, ERC IUT and ERDF (Estonia); Academy of Finland, MEC, and HIP (Finland); CEA and CNRS/IN2P3 (France); BMBF, DFG, and HGF (Germany); GSRT (Greece); OTKA and NIH (Hungary); DAE and DST (India); IPM (Iran); SFI (Ireland); INFN (Italy); MSIP and NRF (Republic of Korea); LAS (Lithuania); MOE and UM (Malaysia); CINVESTAV, CONACYT, SEP, and UASLP-FAI (Mexico); MBIE (New Zealand); PAEC (Pakistan); MSHE and NSC (Poland); FCT (Portugal); JINR (Dubna); MON, RosAtom, RAS and RFBR (Russia); MESTD (Serbia); SEIDI and CPAN (Spain); Swiss Funding Agencies (Switzerland); MST (Taipei); ThEPCenter, IPST, STAR and NSTDA (Thailand); TUBITAK and TAEK (Turkey); NASU and SFFR (Ukraine); STFC (United Kingdom); DOE and NSF (USA).

Individuals have received support from the Marie-Curie programme and the European Research Council and EPLANET (European Union); the Leventis Foundation; the A. P. Sloan Foundation; the Alexander von Humboldt Foundation; the Belgian Federal Science Policy Office; the Fonds pour la Formation \`a la Recherche dans l'Industrie et dans l'Agriculture (FRIA-Belgium); the Agentschap voor Innovatie door Wetenschap en Technologie (IWT-Belgium); the Ministry of Education, Youth and Sports (MEYS) of the Czech Republic; the Council of Science and Industrial Research, India; the HOMING PLUS programme of the Foundation for Polish Science, cofinanced from European Union, Regional Development Fund; the OPUS programme of the National Science Center (Poland); the Compagnia di San Paolo (Torino); MIUR project 20108T4XTM (Italy); the Thalis and Aristeia programmes cofinanced by EU-ESF and the Greek NSRF; the National Priorities Research Program by Qatar National Research Fund; the Rachadapisek Sompot Fund for Postdoctoral Fellowship, Chulalongkorn University (Thailand); the Chulalongkorn Academic into Its 2nd Century Project Advancement Project (Thailand); and the Welch Foundation, contract C-1845.
\end{acknowledgments}

\bibliography{auto_generated}

\ifthenelse{\boolean{cms@external}}{}{
\clearpage
\appendix
\section{Supplemental material\label{app:supp-mat}}
\begin{table*}[ht]
\centering
\topcaption{Signal yields, $N$, and background fractions, \fbg (in~\%),
within $1\sigma$ windows around the nominal \UpsN masses
for the considered \cpm and \pt ranges.}
\label{tab:NumEventsFracBG}
\begin{tabular}{cccccccc}
\hline
$\pt$ & \multirow{2}{*}{\cpm} &
\multicolumn{2}{c}{\UpsOne} &
\multicolumn{2}{c}{\UpsTwo} &
\multicolumn{2}{c}{\UpsThree} \\
\,[\GeVns{}] & &
$N$ & \fbg &
$N$ & \fbg &
$N$ & \fbg \\
\hline
\rule{0pt}{4mm}
&  0--10   &  26\,900    & 3    &  10\,200  &    7  &    6\,500 &  10 \\
& 10--20   &  51\,300    & 5    &  18\,200  &   12  &  10\,800 &  18 \\
10--15
& 20--30   &  39\,200    & 7    &  13\,000  &   18  &    7\,500 &  26 \\
& 30--40   &  20\,100    & 9    &   6\,300  &   24  &
\multirow{2}{*}{5\,200} & \multirow{2}{*}{40} \\
& 40--60   &  11\,300    & 12  &    3\,400  &  31  &   & \\ \hline
\rule{0pt}{4mm}
&  0--10   &  10\,600    & 3    &    4\,500  &    7  &  3\,200 &  9 \\
& 10--20   &  27\,900    & 4    &   12\,100  &    9  &  8\,200 &  12 \\
15--35
& 20--30   &  25\,700    & 5    &   10\,900  &  11   &  7\,200 &  15 \\
& 30--40   &  15\,100    & 6    &    6\,300  &  14   &  3\,800 &  20 \\
& 40--60   &  9\,500     & 7    &    3\,400  &  17   &  2\,300 &  24 \\
\hline
\end{tabular}
\end{table*}

\begin{table*}[h!tb]
\centering
\topcaption{\UpsN polarization parameters in the HX frame. The global uncertainties, independent of state and \cpm\ bin, are also indicated.}
\vglue2mm
\begin{tabular}{ccccccc}
\hline\noalign{\smallskip}
State & $\pt$\;[\GeVns{} & \cpm & $\lambda_\vartheta$ & $\lambda_\varphi$ & $\lambda_{\vartheta\varphi}$ & $\tilde\lambda$ \\ \noalign{\smallskip}
\hline\noalign{\smallskip}
\multirow{5}{*}{$\Upsilon$(1S)} & \multirow{5}{*}{10--15}
    & 0--10  & $+0.014^{+0.102}_{-0.110} $ & $-0.044^{+0.026}_{-0.029} $ & $+0.053^{+0.033}_{-0.036} $ & $-0.094^{+0.126}_{-0.138} $ \\ \noalign{\smallskip}
& &10--20  & $+0.103^{+0.089}_{-0.098} $ & $-0.035^{+0.025}_{-0.027} $ & $+0.027^{+0.028}_{-0.031} $ & $+0.005^{+0.126}_{-0.139} $ \\ \noalign{\smallskip}
& & 20--30 & $+0.187^{+0.102}_{-0.110} $ & $-0.041^{+0.026}_{-0.029} $ & $+0.035^{+0.033}_{-0.037} $ & $+0.066^{+0.128}_{-0.145} $ \\ \noalign{\smallskip}
& & 30--40 & $+0.032^{+0.121}_{-0.132} $ & $-0.042^{+0.027}_{-0.031} $ & $+0.033^{+0.039}_{-0.044} $ & $-0.074^{+0.146}_{-0.159} $ \\ \noalign{\smallskip}
& & 40--60 & $-0.073^{+0.159}_{-0.174} $ & $-0.074^{+0.034}_{-0.037} $ & $+0.061^{+0.054}_{-0.059} $ & $-0.243^{+0.186}_{-0.201} $ \\ \noalign{\smallskip}
\hline\noalign{\smallskip}
\multirow{5}{*}{$\Upsilon$(2S)} & \multirow{5}{*}{10--15}
&  0--10  & $+0.232^{+0.140}_{-0.156} $ & $-0.018^{+0.035}_{-0.037} $ & $-0.048^{+0.049}_{-0.054} $ & $+0.166^{+0.174}_{-0.183} $ \\ \noalign{\smallskip}
& &10--20  & $+0.214^{+0.126}_{-0.136} $ & $-0.047^{+0.032}_{-0.035} $ & $-0.055^{+0.042}_{-0.048} $ & $+0.075^{+0.158}_{-0.178} $ \\ \noalign{\smallskip}
& &20--30  & $+0.230^{+0.158}_{-0.171} $ & $-0.028^{+0.036}_{-0.040} $ & $-0.045^{+0.057}_{-0.061} $ & $+0.137^{+0.201}_{-0.222} $ \\ \noalign{\smallskip}
& &30--40  & $+0.267^{+0.213}_{-0.226} $ & $-0.024^{+0.044}_{-0.046} $ & $-0.059^{+0.072}_{-0.084} $ & $+0.183^{+0.272}_{-0.284} $ \\ \noalign{\smallskip}
& &40--60  & $+0.317^{+0.289}_{-0.325} $ & $-0.061^{+0.055}_{-0.063} $ & $+0.023^{+0.103}_{-0.113} $ & $+0.128^{+0.345}_{-0.366} $ \\ \noalign{\smallskip}
\hline\noalign{\smallskip}
\multirow{4}{*}{$\Upsilon$(3S)} & \multirow{4}{*}{10--15}
&  0--10  & $+0.074^{+0.169}_{-0.187} $ & $-0.028^{+0.038}_{-0.044} $ & $-0.044^{+0.061}_{-0.069} $ & $-0.002^{+0.205}_{-0.219} $ \\ \noalign{\smallskip}
& &10--20  & $+0.279^{+0.156}_{-0.173} $ & $-0.019^{+0.037}_{-0.040} $ & $-0.148^{+0.060}_{-0.064} $ & $+0.208^{+0.209}_{-0.223} $ \\ \noalign{\smallskip}
& &20--30  & $+0.061^{+0.194}_{-0.213} $ & $-0.032^{+0.043}_{-0.047} $ & $+0.061^{+0.069}_{-0.078} $ & $-0.023^{+0.245}_{-0.259} $ \\ \noalign{\smallskip}
& &30--60  & $+0.672^{+0.285}_{-0.304} $ & $+0.051^{+0.061}_{-0.068} $ & $-0.196^{+0.108}_{-0.116} $ & $+0.798^{+0.435}_{-0.479} $ \\ \noalign{\smallskip}
\hline\noalign{\smallskip}
& & global unc.\ & $\pm 0.085$ & $\pm 0.023$ & $\pm 0.022$ & $\pm 0.104$ \\ \noalign{\smallskip}
\hline\noalign{\smallskip}
\multirow{5}{*}{$\Upsilon$(1S)} & \multirow{5}{*}{15--35}
 & 0--10  & $-0.002^{+0.102}_{-0.110} $ & $-0.054^{+0.029}_{-0.032} $ & $+0.020^{+0.038}_{-0.041} $ & $-0.142^{+0.128}_{-0.138} $ \\ \noalign{\smallskip}
& &10--20  & $-0.004^{+0.087}_{-0.094} $ & $-0.047^{+0.037}_{-0.041} $ & $-0.000^{+0.033}_{-0.038} $ & $-0.127^{+0.130}_{-0.142} $ \\ \noalign{\smallskip}
& &20--30  & $+0.053^{+0.087}_{-0.093} $ & $-0.054^{+0.039}_{-0.042} $ & $-0.005^{+0.034}_{-0.037} $ & $-0.095^{+0.121}_{-0.130} $ \\ \noalign{\smallskip}
& &30--40  & $+0.036^{+0.097}_{-0.105} $ & $-0.026^{+0.033}_{-0.036} $ & $-0.000^{+0.037}_{-0.042} $ & $-0.037^{+0.128}_{-0.139} $ \\ \noalign{\smallskip}
& &40--60  & $-0.021^{+0.106}_{-0.112} $ & $-0.049^{+0.039}_{-0.044} $ & $+0.007^{+0.043}_{-0.046} $ & $-0.148^{+0.151}_{-0.164} $ \\ \noalign{\smallskip}
\hline\noalign{\smallskip}
\multirow{5}{*}{$\Upsilon$(2S)} & \multirow{5}{*}{15--35}
 & 0--10  & $+0.124^{+0.163}_{-0.168} $ & $-0.058^{+0.034}_{-0.039} $ & $-0.050^{+0.048}_{-0.054} $ & $-0.039^{+0.189}_{-0.190} $ \\ \noalign{\smallskip}
& &10--20  & $+0.237^{+0.105}_{-0.114} $ & $-0.027^{+0.028}_{-0.032} $ & $-0.025^{+0.039}_{-0.043} $ & $+0.149^{+0.138}_{-0.148} $ \\ \noalign{\smallskip}
& &20--30  & $+0.205^{+0.104}_{-0.113} $ & $-0.086^{+0.030}_{-0.034} $ & $+0.052^{+0.038}_{-0.043} $ & $-0.040^{+0.141}_{-0.146} $ \\ \noalign{\smallskip}
& &30--40  & $+0.501^{+0.147}_{-0.153} $ & $-0.020^{+0.033}_{-0.037} $ & $-0.078^{+0.047}_{-0.053} $ & $+0.416^{+0.177}_{-0.184} $ \\ \noalign{\smallskip}
& &40--60  & $+0.364^{+0.171}_{-0.172} $ & $-0.061^{+0.040}_{-0.044} $ & $-0.024^{+0.053}_{-0.061} $ & $+0.169^{+0.200}_{-0.206} $ \\ \noalign{\smallskip}
\hline\noalign{\smallskip}
\multirow{5}{*}{$\Upsilon$(3S)} & \multirow{5}{*}{15--35}
 & 0--10  & $+0.381^{+0.188}_{-0.193} $ & $-0.074^{+0.039}_{-0.044} $ & $+0.048^{+0.059}_{-0.066} $ & $+0.144^{+0.210}_{-0.203} $ \\ \noalign{\smallskip}
& &10--20  & $+0.244^{+0.121}_{-0.134} $ & $-0.029^{+0.033}_{-0.037} $ & $-0.035^{+0.041}_{-0.048} $ & $+0.151^{+0.169}_{-0.177} $ \\ \noalign{\smallskip}
& &20--30  & $+0.285^{+0.121}_{-0.132} $ & $-0.048^{+0.032}_{-0.036} $ & $-0.053^{+0.041}_{-0.048} $ & $+0.135^{+0.162}_{-0.165} $ \\ \noalign{\smallskip}
& &30--40  & $+0.220^{+0.163}_{-0.169} $ & $-0.121^{+0.039}_{-0.043} $ & $-0.050^{+0.054}_{-0.060} $ & $-0.114^{+0.191}_{-0.197} $ \\ \noalign{\smallskip}
& &40--60  & $+0.278^{+0.218}_{-0.206} $ & $-0.072^{+0.048}_{-0.051} $ & $-0.143^{+0.068}_{-0.078} $ & $+0.057^{+0.246}_{-0.231} $ \\ \noalign{\smallskip}
\hline\noalign{\smallskip}
& & global unc.\ & $\pm 0.066$ & $\pm 0.019$ & $\pm 0.020$ & $\pm 0.073$ \\ \noalign{\smallskip}
\hline\noalign{\smallskip}
\end{tabular}
\end{table*}

}
\cleardoublepage \section{The CMS Collaboration \label{app:collab}}\begin{sloppypar}\hyphenpenalty=5000\widowpenalty=500\clubpenalty=5000\textbf{Yerevan Physics Institute,  Yerevan,  Armenia}\\*[0pt]
V.~Khachatryan, A.M.~Sirunyan, A.~Tumasyan
\vskip\cmsinstskip
\textbf{Institut f\"{u}r Hochenergiephysik der OeAW,  Wien,  Austria}\\*[0pt]
W.~Adam, E.~Asilar, T.~Bergauer, J.~Brandstetter, E.~Brondolin, M.~Dragicevic, J.~Er\"{o}, C.~Fabjan\cmsAuthorMark{1}, M.~Flechl, M.~Friedl, R.~Fr\"{u}hwirth\cmsAuthorMark{1}, V.M.~Ghete, C.~Hartl, N.~H\"{o}rmann, J.~Hrubec, M.~Jeitler\cmsAuthorMark{1}, V.~Kn\"{u}nz, A.~K\"{o}nig, M.~Krammer\cmsAuthorMark{1}, I.~Kr\"{a}tschmer, D.~Liko, T.~Matsushita, I.~Mikulec, D.~Rabady\cmsAuthorMark{2}, N.~Rad, B.~Rahbaran, H.~Rohringer, J.~Schieck\cmsAuthorMark{1}, R.~Sch\"{o}fbeck, J.~Strauss, W.~Treberer-Treberspurg, W.~Waltenberger, C.-E.~Wulz\cmsAuthorMark{1}
% ...
\vskip\cmsinstskip
% ...
1:~~Also at Vienna University of Technology, Vienna, Austria\\
2:~~Also at CERN, European Organization for Nuclear Research, Geneva, Switzerland\\
% ...

\end{sloppypar}
\end{document}